\documentstyle[preprint,aps,prd,eqsecnum,graphicx]{revtex}

\begin{document}
\draft


\title{\vspace*{1\baselineskip}
         Cosmological constant and vacuum energy}

\author{G. E. Volovik\footnote{Email address:
             volovik@boojum.hut.fi}}
\address{Helsinki University of Technology,
Low Temperature Laboratory,\\ P.O. Box 2200, FIN--02015 HUT,  Finland, \\
Landau Institute for Theoretical Physics, 119334 Moscow, Russia}

\date{October 18, 2004}   

\maketitle

\begin{abstract}
  The general thermodynamic analysis of the quantum vacuum, which 
is based on our knowledge of the vacua in condensed-matter systems, is
consistent with the Einstein earlier view on the cosmological constant.
In the equilibrium Universes the value of the cosmological constant is
regulated by matter. In the empty Universe, the vacuum energy is exactly
zero, $\lambda=0$. The huge contribution of the zero point motion of the
quantum fields to the vacuum energy is exactly cancelled by the
higher-energy degrees of freedom of the quantum vacuum.  In the
equilibrium Universes homogeneously filled by matter, the vacuum is
disturbed, and the energy density of the vacuum becomes proportional to that of matter,  $\lambda=\rho_{\rm vac}\sim \rho_{\rm
matter}$. This consideration applies to any  vacuum in equilibrium
irrespective of whether the vacuum is false or true, and is valid both in
Einstein's general theory of relativity and within the special theory
of relativity, i.e. in a world without gravity. 

\end{abstract}
\maketitle                   





\section{Introduction}

In 1917, Einstein proposed a model of our Universe \cite{einstein}. 
To make the Universe
static, he introduced the famous cosmological constant which 
was counterbalancing the collapsing tendency of the gravitating matter.
 As a static solution 
of the field equations of general relativity with added cosmological term, he obtained the Universe with spatial geometry of a three-dimensional sphere. In Einstein treatment the cosmological
constant is universal, i.e. it must be constant throughout the whole
Universe. But it is not fundamental: its value is determined by the
matter density in the Universe.  In Ref. \cite{bib1}, Einstein noted 
that the $\lambda$-term must be added to
his equations if the density of matter in the Universe is non-zero 
in average.
 In particular, this means that $\lambda=0$ if matter in the Universe
is so inhomogeneously distributed that its average over big volumes $V$ tends to zero. 
In this treatment, $\lambda$ resembles a Lagrange multiplier or
an integration constant,  rather than the fundamental constant (see general
discussion in Ref. \cite{Weinberg2,Padmanabhan}).

With the development of the quantum field theory it
was recognized that the $\lambda$-term is related to zero-point motion 
of quantum fields. It describes
the energy--momentum tensor of the quantum vacuum, 
$T^{\mu\nu}_{\rm vac} = \lambda g^{\mu\nu}$. This means that $\lambda$ is
nothing but the energy density of the vacuum, 
$\lambda=\rho_{\rm vac}$, i.e. the vacuum can be 
considered as a medium obeying the
equation of state: 
\begin{equation}
\rho_{\rm vac}=-p_{\rm vac}~.
\label{VacuumEOS}
\end{equation} 
Such view on the cosmological constant led to principle
difficulties. The main two problems are: (i)  the energy density of the
zero-point motion is highly divergent because of the formally
infinite number of modes; (ii) the vacuum energy is determined by the
high-energy degrees of quantum fields, and thus at first glance must
have a fixed value which is not sensitive to the low-density and
low-energy matter in the present Universe, which is also in disagreement 
with observations. 

The naive summation over all the known modes of the
quantum fields gives the following estimate for the energy density
of the quantum vacuum  
\begin{equation}
\rho_{\rm vac}= {1\over V}\left(  {1\over 2} \sum_b \sum_{\bf
p}  E_b(p)~~
- \sum_f
\sum_{\bf p} E_f(p)\right)~.
\label{VacuumEnergyPlanck1}
\end{equation} 
Here the negative contribution comes from the negative energy
levels occupied by fermionic species $f$ in the Dirac sea; the
positive contribution comes from the zero-point energy of quantum
fluctuations of bosonic fields $b$.  Since the largest contribution
comes from the quantum fluctuations with ultrarelativistic momenta
$p\gg mc$, the masses $m$ of particles can be neglected, and the
energy spectrum of particles can be considered as massless,
$E_b(p)=E_f(p)=cp$. Then the energy density of the quantum vacuum is
expressed in terms of the number $\nu_b$ and $\nu_f$ of 
bosonic and fermionic species:
\begin{equation}
\rho_{\rm vac} = {1\over V}\left({1\over 2}\nu_b\sum_{\bf
p} cp~~ -\nu_f
\sum_{\bf p} cp\right) \sim {1\over c^3} \left({1\over 2}\nu_b 
-\nu_f\right) E_{\rm Pl}^4~.
\label{VacuumEnergyPlanck2}
\end{equation} 
Here $E_{\rm Pl}$ is the Planck energy cut-off.  This
estimate of the cosmological constant exceeds by 120 orders of
magnitude the upper limit posed by astronomical observations. The
more elaborated calculations of the vacuum energy, which take into
account the interaction between different modes in the vacuum, can
somewhat reduce the estimate but not by many orders of
magnitude. The supersymmetry  --
the symmetry between the fermions and bosons which imposes
the relation $\nu_b=2\nu_f$ -- does not help too. In our world the
supersymmetry is not exact, and one obtains
$\rho_{\rm vac}\sim {1\over c^3}   E_{\rm UV}^4$, where the ultra-violet
cut-off $E_{\rm UV}$ is provided by the energy scale below which the
supersymmetry is violated. If it exists, the supersymmetry can
substantially reduce this estimate, but still a discrepancy remains of at least 60 orders of
magnitude.

Moreover, the cut-off energy is the intrinsic parameter of 
quantum field theory. It is determined by the high energy
degrees of freedom of the order of $E_{\rm UV}$ or $E_{\rm Pl}$ and 
thus cannot be
sensitive to the density of matter in the present Universe. The
typical energies of the present matter are too low compared to 
$E_{\rm UV}$, and thus the matter is unable to influence such a deep
structure of the vacuum. This contradicts to recent observations which
actually support the Einstein prediction that the cosmological
constant is determined by the energy density of matter
$\rho_{\rm matter}$. At the moment the consensus has emerged about the
experimental value of the cosmological constant
\cite{Spergel,Padmanabhan}. It is on the order of magnitude of the matter
density, $\rho_{\rm vac}\sim 2-3\rho_{\rm matter}$. This is comparable to 
\begin{equation}
\rho_{\rm vac}={1\over 2}\rho_{\rm matter} ~
\label{VacuumMatterEinstein}
\end{equation} 
obtained by Einstein for the static cold Universe, and
\begin{equation}
\rho_{\rm vac}= \rho_{\rm matter} ~
\label{VacuumMatterEinstein2}
\end{equation} 
in the static hot Universe filled by ultra-relativistic matter or
 radiation
(see Eq. (\ref{EinsteinSolution}) below). 

The pressure of the vacuum was found to be negative, 
$p_{\rm vac}=-\rho_{\rm vac}<0$, which means that the vacuum  does
really oppose and partially counterbalance the collapsing tendency
of matter.  This demonstrates that, though our Universe is expanding
(even with acceleration) and is spatially flat, it is not very far from the 
Einstein's static equilibrium solution. 

The problem is how to reconcile the astronomical observations with
the estimate of the vacuum energy imposed by the relativistic
Quantum Field Theory (QFT). What is the flaw in the arguments which
led us to Eq. (\ref{VacuumEnergyPlanck2}) for the vacuum energy? The
evident weak point is that the summation over the modes 
in the quantum vacuum is
constrained by the cut-off: we are not able to sum over all degrees
of freedom of the quantum vacuum since we do not know the physics of
the deep vacuum beyond the cut-off. It is quite possible that we
simply are not aware of some very simple principles of the
trans-Planckian physics from which it immediately follows that the
correct summation over all the modes of the quantum vacuum
gives zero or almost zero value for the vacuum energy density, i.e.
the trans-Planckian degrees of freedom effectively cancel the
contribution of the sub-Planckian degrees irrespective of details of
trans- and sub-Planckian physics. People find it easier to believe that
such an unknown mechanism of cancellation if it existed would
 reduce $\lambda$ to exactly zero rather than the observed very low value.

Since we are looking for the general principles governing  the energy
of the vacuum, it should not be of importance for us whether the QFT
is fundamental or emergent. Moreover, we expect that these principles
should not depend on whether or not  the QFT obeys all the
symmetries of the relativistic QFT: these symmetries (Lorentz and
gauge invariance, supersymmetry, etc.) still did not help us to
nullify the vacuum energy).  That is why to find these principles we
can look at the quantum vacua whose microscopic structure is well
known at least in principle. These are the ground states of the
quantum condensed-matter systems, such as superfluid liquids,
Bose-Einstein condensates in ultra-cold gases, superconductors, insulators, systems
experiencing the quantum Hall effect, etc. These systems provide us
with a broad class of Quantum Field Theories  which are not
restricted by Lorentz invariance. This allows us to consider many
problems in the relativistic Quantum Field Theory of the
weak, strong and electromagnetic interactions and gravitation from a
more general perspective. In particular, the cosmological constant
problems:  Why is $\lambda$ not big? Why is it non-zero? Why is it of the
order of magnitude of the matter density? ...

\section{Effective QFT in quantum liquids} 

The homogeneous ground state of a quantum system, even though it contains
a large amount of particles (atoms or electrons), does really play
the role of a quantum vacuum. Quasiparticles -- the
propagating low-frequency excitations above the ground state, 
that play the role of elementary particles in the effective QFT -- 
see the ground state as an empty space. For example,  phonons -- 
the quanta of the sound 
waves in
superfluids -- do not scatter on the atoms of the liquid if the atoms
are in their ground state. The interacting bosonic and fermionic 
quasiparticles
are described by the bosonic and fermionic quantum fields, obeying the
same principles of the QFT except that in general they are not
relativistic 
and do not obey the symmetries of relativistic QFT. 
They obey at most the Galilean 
invariance and have a preferred reference frame where 
the liquid is at rest. It is
known, however, that in some of these systems 
the effective Lorentz symmetry emerges for quasiparticles. Moreover, if 
the system belongs to a special universality class, the Lorentz
symmetry emerges together with effective gauge and metric fields 
\cite{VolovikBook}. This fact, though encouraging for other
applications of condensed matter methods to relativistic QFT (see
e.g. \cite{KlinkhamerVolovik}), is not important for our
consideration. The principle which leads to nullification of
the vacuum energy is more general, it comes from a thermodynamic 
analysis
which is not constrained by symmetry or universality class. 

To see it let us consider two quantum vacua: the ground states of two
quantum liquids, superfluid $^4$He and one of the two superfluid
phases of  $^3$He, the A-phase. We have chosen these two liquids
because the spectrum of quasiparticles playing the major role at
low energy is `relativistic', i.e. $E(p)=cp$, where $c$ is some
parameter of the system. This allows us to make the connection to
relativistic QFT. In superfluid $^4$He the relevant  quasiparticles 
are phonons (quanta of sound waves), and $c$ is the speed of sound.
In superfluid $^3$He-A  the relevant  quasiparticles  are fermions. 
The corresponding `speed of light' $c$ (the slope in the linear spectrum
of these fermions) is anisotropic; it depends on the direction of their
propagation: $E^2({\bf p})=c_x^2p_x^2+c_y^2p_y^2+c_z^2p_z^2$. But this
detail is not important for our consideration. 

According to the naive
estimate in Eq. (\ref{VacuumEnergyPlanck2}) the density of the
ground state energy in the bosonic liquid $^4$He comes from the
zero-point motion of the phonons
\begin{equation}
\rho_{\rm vac}={1\over 2} \sum_{\bf
p}  cp \sim {E_{\rm Pl}^4\over c^3}= E_{\rm Pl}^4\sqrt{-g}  ~,
\label{4He}
\end{equation} 
where the ultraviolet cut-off is provided by the Debye temperature,
$E_{\rm Pl}= E_{\rm Debye}\sim 1$ K; $c\sim 10^4$ cm/s; and we
introduced the
 effective acoustic metric for phonons \cite{UnruhSonic}. The ground
state energy of  fermionic liquid must come from the occupied
negative energy levels of the Dirac sea:
 \begin{equation}
\rho_{\rm vac}  \sim 
-{ E_{\rm Pl}^4\over c_x c_y c_z} =-E_{\rm Pl}^4\sqrt{-g}~.
\label{3He}
\end{equation} 
Here the `Planck' cut-off is provided by the amplitude of the
superfluid order parameter, $E_{\rm Pl}= \Delta \sim 1$ mK; 
$c_z\sim 10^4$ cm/s;
$c_x=c_y \sim 10$ cm/s.

These estimates were obtained by using the effective QFT for the 
`relativistic'
fields.
Comparing them  with the results obtained by using the known microscopic
physics of these liquids one finds that these estimates are not
completely crazy: they do reflect some important part of
microscopic physics. For example, the Eq. (\ref{3He}) gives the
correct order of magnitude for the difference between the energy
densities of the liquid $^3$He in superfluid state, which represents
the true vacuum, and the normal (non-superfluid) state  representing
the false vacuum: 
\begin{equation}
\rho_{\rm true}- \rho_{\rm false} \sim -{ E_{\rm Pl}^4\over c_x c_y
c_z} ~.
\label{3He2}
\end{equation} 
However, it says nothing on the total energy density of the liquid.
 Moreover, as we shall see below, it also gives a disparity of many orders of magnitude  between the estimated and measured values of the
analog of the cosmological constant in this liquid. Thus in the 
condensed-matter vacua we have the same paradox with the vacuum energy.
But the advantage is that we know the microscopic physics of the quantum
vacuum in these systems and thus are able to resolve the paradox there.

\section{Relevant thermodynamic potential for quantum vacuum} 

When one discusses the energy of condensed matter, one must
specify what thermodynamic potential is relevant for the particular
problem which he or she considers. Here we are interested in the
analog of the QFT emerging in condensed matter.  The many-body system of
the collection of  identical atoms 
(or electrons) obeying Schr\"odinger quantum mechanics can be described in
terms of the QFT
\cite{AGDbook}
 whose Hamiltonian is
\begin{equation}
{\cal H}-\sum_a \mu_a{\cal N}_a~.
\label{Hamiltonian}
\end{equation}
 Here ${\cal H}$ is the second-quantized Hamiltonian of the
 many-body system containing the fixed numbers of atoms of different
sorts. It is expressed in terms of the Fermi and Bose quantum  fields
$\psi_a({\bf} r, t)$.  The operator
${\cal N}_a=\int d^3  \psi^\dagger_a\psi_a$ is the particle number
 operator for atoms of sort $a$.   The Hamiltonian (\ref{Hamiltonian}) removes the
constraint imposed on the quantum fields $\psi_a$ by the conservation law
for the number of atoms of sort $a$, and it corresponds to the thermodynamic
potential with fixed chemical potentials $\mu_a$.The Hamiltonian (\ref{Hamiltonian}) also
serves as a starting point for the construction of the effective QFT for
quasiparticles, and thus it is responsible for their vacuum. 

Thus the correct vacuum energy density for the QFT emerging in the
many-body system
 is 
determined by the vacuum expectation value of fhe Hamiltonian 
(\ref{Hamiltonian}) in the thermodynamic limit $V\rightarrow \infty$ and
$N_a\rightarrow \infty$:
\begin{equation}
\rho_{\rm
vac}  ={1\over V}\left<{\cal H}-\sum_a \mu_a{\cal
N}_a\right>_{\rm vac}~.
\label{VacuumEnergy}
\end{equation}

One can check that this is the right choice for
the vacuum energy using the Gibbs-Duhem relation of
thermodynamics. It states that if the condensed matter is in
equilibrium it obeys the following relation between the energy
$E=\left<{\cal H}\right>$, and  the other thermodynamic varaibles --
the temperature  $T$, the entropy
$S$, the particle numbers $N_a=\left<{\cal N}_a\right>$, the
chemical potentials $\mu_a$, and the pressure $p$: 
\begin{equation}
E-TS-\sum_a \mu_a  N_a =-pV~.
\label{Gibbs-DuhemRelation}
\end{equation}
Applying this thermodynamic Gibbs-Duhem relation to the ground state
at $T=0$ and using Eq. (\ref{VacuumEnergy}) one obtains  
\begin{equation}
\rho_{\rm vac}  ={1\over V} \left(E-\sum_a \mu_a
N_a\right)_{\rm vac}=-p_{\rm vac} ~.
\label{EnergyPressure}
\end{equation}
Omitting the intermediate expression, the second term in
Eq. (\ref{EnergyPressure}) which contains  the microscopic parameters $\mu_a$,
one finds the familiar equation of state for the vacuum -- 
the equation (\ref{VacuumEOS}).  The vacuum is a medium with
 the equation of state 
(\ref{VacuumEOS}), and such a medium naturally emerges in any 
condensed-matter QFT, relativistic or non-relativistic. 
This demonstrates that the problem
of the vacuum energy can be considered from the more general
perspective not constrained by the relativistic Hamiltonians. Moreover, 
it is not important whether there is gravity or not. We shall see below
in Sec. \ref{non-gravitatingUniverse} that the vacuum plays an important
role even in the absence of gravity: it stabilizes the Universe filled
with hot non-gravitating matter

There is one lesson from the microscopic consideration of the vacuum 
energy, which we can learn immediately and which is very important for
one of the problems related to the cosmological constant. The problem is that
while on the one hand the physical laws do not change if we add a constant  to
the Hamiltonian, i.e. they are invariant under the transformation ${\cal H}
\rightarrow {\cal H}+C$, on the other hand gravity responds to the whole
energy and thus is sensitive to the choice of $C$.  Our condensed-matter
QFT shows how this problem  can be resolved. Let us shift the energy of
each atom of the many-body system by the same amount $\alpha$. This
certainly changes the original many-body Hamiltonian ${\cal H}$ for the
fixed number of atoms: after the shift it becomes ${\cal H}+\alpha \sum_a
N_a$. But the proper Hamiltonian (\ref{Hamiltonian}), which is relevant
for the QFT, remains invariant under this transformation, ${\cal H}
-\sum_a \mu_a{\cal N}_a\rightarrow{\cal H} -\sum_a \mu_a{\cal N}_a$. This
is because the chemical potentials are also shifted: $\mu_a \rightarrow
\mu_a+\alpha $.

This demonstrates that  when the proper thermodynamic potential is used, 
the vacuum energy becomes independent of the choice of the reference for
the energy. This is the general thermodynamic property which does not
depend on details of the many-body system. This suggests that one of the
puzzles of the cosmological constant  --  that $\lambda$ depends on the
choice of zero energy level -- could simply result from our very limited
knowledge of the quantum vacuum. We are unable to see the robustness of
the vacuum energy from our low-energy corner, we need a deeper
thermodynamic analysis. But the result of this analysis does not depend
on the details of the structure of the quantum vacuum. In particular, it does not
depend on how many different chemical potentials $\mu_a$ are at the 
microscopic level: one, several or none. That is why we expect that 
this general thermodynamic analysis could be applied to our vacuum too.

\section{Nullification of vacuum energy in the equilibrium vacuum} 

Now let us return to our two monoatomic quantum liquids, $^3$He and
$^4$He, each with a single chemical potential $\mu$, and calculate the
relevant ground-state energy (\ref{VacuumEnergy}) in each of them.  Let us
consider the simplest situation, when our liquids are completely isolated
from the environment. For example, one can consider the  quantum liquid
in space where it forms a droplet.  
 Let us assume that the radius $R$ of the droplet is so big that we 
can neglect the contribution of the surface effects to the energy
density. The evaporation at $T=0$ is absent, that is why the ground state
exists and we can calculate its energy from the first principles. Though
both liquids  are collections of strongly interacting and strongly
correlated atoms, numerical simulations of the ground state energy have
been done with a very simple result. In the limit $R\rightarrow \infty$
and $T=0$ the energy density of both liquids $\rho_{\rm vac}\rightarrow
0$.   The zero result is in apparent contradiction with Eqs. (\ref{4He})
and  (\ref{3He}). But it is not totally unexpected since it is in
complete agreement with Eq. (\ref{EnergyPressure}) which follows from the
Gibbs-Duhem relation: in the absence of external environment the external
pressure is zero, and thus the pressure of the liquid in its equilibrium
ground state
$p_{\rm vac}=0$. Therefore $\rho_{\rm vac}  =-p_{\rm vac}=0$, and this
nullification occurs irrespective of whether the liquid is made of
fermionic or bosonic atoms.

If the observers living within the droplet measure the vacuum energy (or
the vacuum pressure) and compare it with their estimate, Eq. (\ref{4He})
or  Eq. (\ref{3He}) depending on in which liquid they live, they will be
surprised by the disparity of many orders of magnitude between the
estimate and observation. But we can easily explain to these observers
where the mistake is.  The equations (\ref{4He}) and  (\ref{3He}) take
into account only the degrees of freedom below the cut-off energy. If one
takes into account all the degrees of freedom, not only the low-energy
modes of the effective QFT, but the real microscopic energy of
interacting atoms (what the low-energy observer is unable to do), the
zero result will be obtained. The exact cancellation occurs without any
special fine-tuning: the microscopic degrees of freedom
of the system perfectly neutralize the huge contribution of the
sub-Planckian modes due to the thermodynamic relation applied to the
whole equilibrium ground state.

The above thermodynamic analysis does not depend on the microscopic 
structure of the vacuum and thus can be applied to any quantum vacuum,
including the vacuum of relativistic QFT. This is another lesson from
condensed matter which we may or may not accept: the energy density of
the homogeneous equilibrium state of the quantum vacuum is zero in the
absence of external environment. The higher-energy 
(trans-Planckian) degrees of freedom of the quantum vacuum perfectly
cancel the huge contribution of the zero-point motion of the quantum
fields to the vacuum energy. This occurs without fine-tuning because of
the underlying general thermodynamic laws.

There exists a rather broad belief that the problem of the vacuum
energy can be avoided simply by the proper choice of the ordering of the
QFT operators $\psi_a$ and $\psi^\dagger_a$.  However, this does not work
in situations when the vacuum is not unique or is perturbed, which we
discuss below. In our quantum liquids, the zero result has been obtained
using the original pre-QFT microscopic theory -- the Schr\"odinger
quantum mechanics of interacting atoms, from which the QFT emerges as a
secondary (second-quantized) theory.  In this approach the problem of the
ordering of the operators in the emergent QFT is resolved on the
microscopic level.

\section{Coincidence problem}
\label{non-gravitatingUniverse}

Let us turn to the second cosmological problem -- the coincidence
problem: Why is in the present Universe the energy density of the
quantum vacuum of the same order of magnitude as the matter density?
To answer this question let us again exploit our quantum liquids as a
guide. Till now we discussed the pure vacuum state, i.e. the state without
matter. In QFT of quantum liquids the matter is represented by
excitations above the vacuum -- quasiparticles. We can introduce thermal
quasiparticles by applying a non-zero temperature $T$ to the liquid
droplets. The quasipartcles in both liquids are `relativistic' and
massless. The pressure of the dilute gas of quasiparticles as a function
of $T$ has the same form in two superfluids if one again uses the
effective metric:
\begin{equation}
p_{\rm matter}=\gamma T^4\sqrt{-g}
 ~. 
 \label{EOSmatter}
\end{equation}
For quasipartcles in $^4$He,
$\sqrt{-g}= c^{-3}$ is the square-root of determinant of the
effective acoustic metric as before, and the parameter $\gamma=\pi^2/90$; 
for
the fermionic quasiparticles in
$^3$He-A, one has $\sqrt{-g}= c_x^{-1}c_y^{-1}c_z^{-1}$ and
$\gamma=7\pi^2/360$. 
Such `relativistic' gas of quasiparticles obeys the ultra-relativistic 
equation of
state for radiation:
\begin{equation}
\rho_{\rm matter}=3p_{\rm matter}
 ~.
\label{EOSmatter}
\end{equation}

Let us consider again the droplet of a quantum liquid which
is isolated from the environment, but now at finite $T$. The new factor
 which is important is the `radiation' pressure produced by the gas of
`relativistic' quasiparticles. In the absence of environment and for
a sufficiently big droplet, when we can neglect the surface tension, the
total pressure in the droplet must be zero.  This means that in
equilibrium, the  partial pressure of matter (quasiparticles) must be
necessarily compensated by the negative pressure of the quantum vacuum
(superfluid condensate):
\begin{equation}
p_{\rm matter}+ p_{\rm vac} =0~.
 \label{ZeroPressure}
\end{equation}
The vacuum pressure leads to vacuum energy density according the equation
of state (\ref{VacuumEOS}) for the vacuum, and one obtains the following
relation between the energy density of the vacuum and that of the
ultra-relativistic matter in the thermodynamic equilibrium:
\begin{equation}
\rho_{\rm vac} =-p_{\rm vac} =p_{\rm matter}={1\over
3}\rho_{\rm matter}~.
\label{VacuumMatterEnergy}
\end{equation} 
This is actually what occurs in quantum liquids, but the resulting
 equation, $\rho_{\rm vac}
={1\over 3}\rho_{\rm matter}$, does not depend on the details of the
system. It is completely determined by the thermodynamic laws and 
equation of state
for matter and is
equally applicable to both quantum systems: (i) superfluid condensate +
quasiparticles with linear  `relativistic'  spectrum; and (ii) vacuum 
 of relativistic quantum fields +
ultra-relativistic matter. That is why we can learn some more lessons from
the condensed-matter examples. 

Let us compare Eq. (\ref{VacuumMatterEnergy}) with Eq. 
(\ref{VacuumMatterEinstein2}) which expresses  the cosmological constant
in terms of the matter density in the 
Einstein Universe also filled with the ultra-relativistic matter. The
difference between them is by a factor 3. The reason is that in the
effective QFT of liquids the Newtonian gravity is absent; the effective
 matter living
in these liquids is non-gravitating: quasiparticles do not
experience the attracting gravitational interaction. Our condensed matter 
reproduces the Universe without gravity, i.e. obeying Einstein's
special theory of relativity. Thus we obtained that even without gravity,
the Universe filled with hot matter can be stabilized by the vacuum, in
this case the negative vacuum pressure counterbalances the expanding
tendency of the hot gas (see also Eq. (\ref{SpecialRelativity}) in Sec.
\ref{EinsteinUniverses} and Ref.  \cite{UniverseSpecialRelativity}).  For
both worlds, with and without gravity, the Einstein prediction in Ref.
\cite{bib1} is correct: the matter homogeneously distributed in space
induces  the non-zero cosmological constant.

This and the other examples lead us to the more general conclusion: when 
the vacuum is disturbed, it responds to perturbation, and the vacuum
energy density becomes non-zero. Applying this to the general relativity,
we conclude that the homogeneous equilibrium state of the quantum vacuum
without matter is not gravitating, but deviations of the quantum vacuum
from such states have weight: they are gravitating. In the above
quantum-liquid examples the vacuum is perturbed by the non-gravitating
matter and also by the surface tension of the curved 2D surface of the
droplet which adds its own partial pressure (see Sec.
\ref{EinsteinUniverses}). In the Einstein Universes it is perturbed by
the gravitating matter and also by the gravitational field (the 3D space
curvature, see Sec. \ref{EinsteinUniverses}). In the expanding or
rotating Universe the vacuum is perturbed by expansion or rotation, etc.
In all these cases, the value of the vacuum energy density is
proportional to the magnitude of perturbations. Since all the
perturbations of the vacuum are small in the present Universe, the
present cosmological constant must be small.  

The special case is when the perturbation (say, matter) occupies a
 finite region of the infinite Universe. In this case the pressure far
outside this region is zero which gives $\lambda=0$. This is in a full
agreement with the statement of Einstein in Ref. \cite{bib1} that the
$\lambda$-term must be added to his equations when the average density of
matter in the Universe is non-zero.

\section{Energy of false and true vacua} 

Let us turn to some other problems related to the cosmological constant. 
For example, what is the energy of the false vacuum and what is the
cosmological constant in such a vacuum? This is important for the
phenomenon of inflation -- the exponential super-luminal expansion of the
Universe. In some theories, the inflation is caused by a false vacuum. It
is usually assumed that the energy of the true vacuum is zero, and thus
the energy of the false vacuum must be positive. Though the false vacuum
can be locally stable at the beginning,   $\lambda$  in this vacuum must
be a big positive constant, which causes the exponential de-Sitter
expansion. Let us look at this scenario using our knowledge of the
general thermodynamic properties of the quantum vacuum. 

Analyzing the Gibbs-Duhem relation we find that in our derivation of 
the vacuum energy, we never used the fact that our system is in the true
ground state. We used only the fact that our system is in the
thermodynamic equilibrium. But this is applicable to the metastable state
too if we neglect the tiny transition processes between the false and
true vacua, such as quantum tunneling  and thermal activation. Thus we
come to the following, at first glance paradoxical, conclusion:
the cosmological constant in all  homogeneous vacua in equilibrium is zero,
irrespective of whether the vacuum is true or false. This poses
constraints on some scenarios of inflation.

\begin{figure} 
\includegraphics[width=\textwidth]{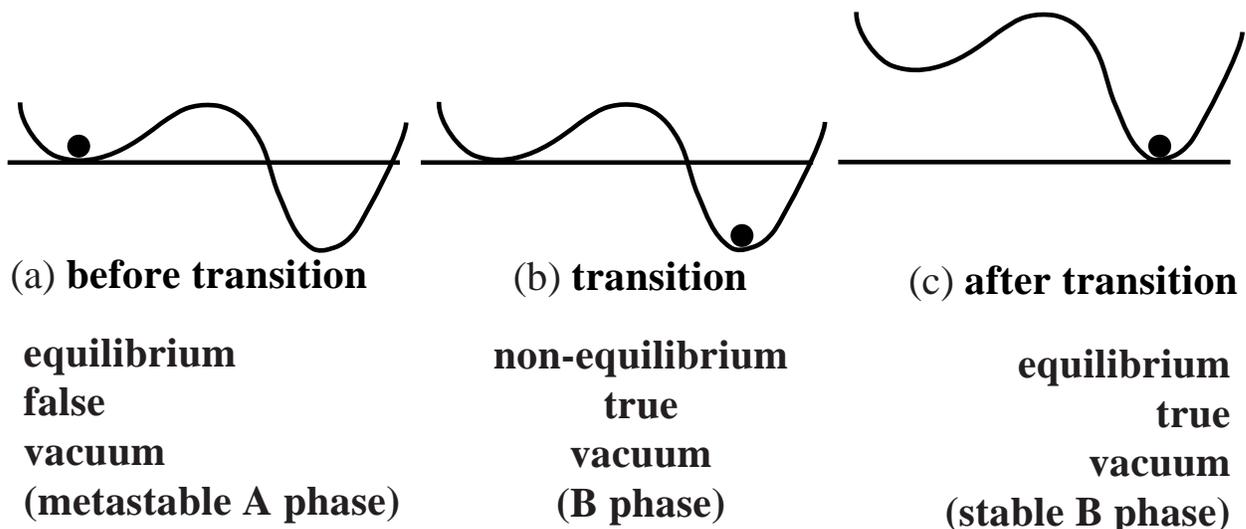}
\caption{ The condensed-matter scenario of the evolution of the energy
density
$\rho_{\rm vac}$ of the quantum vacuum in the process of the first order
phase transition from the equilibrium false vacuum to the equilibrium true
vacuum. Before the phase
transiton, i.e. in the false but equilibrium vacuum, one has $\rho_{\rm
vac}=0$. During the transient period the microscopic parameters of the
vacuum readjust themselves to new equilibrium state, where the equilibrium
condition 
$\rho_{\rm vac}=0$ is restored.}
\label{transition}
\end{figure}

If the vacuum energy is zero both in the false and true vacuum,
then how
and why does the phase transition occur? The thermodynamic analysis for
quantum liquids gives us the answer to this question too. Let us consider
the typical example of the first-order  phase transition which occurs 
between the metastable quantum liquid $^3$He-A and the stable quantum
liquid  $^3$He-B, Fig. \ref{transition}. In the initial metastable but
equilibrium phase A, the thermodynamic potential for this monoatomic
liquid is zero, $E_A-\mu_A N=0$. The same thermodynamic potential
calculated for the phase B at the same $\mu=\mu_A$ is negative:
$E_B-\mu_A N<0$, and thus the liquid prefers the phase transition from
the phase A to phase B. When the transition to the B-phase
occurs, the vacuum energy becomes negative, which corresponds to the
non-equilibrium state. During some transient period of relaxation towards
the thermodynamic equilibrium, the parameter
$\mu$ is readjusted to a new equilibrium state. After that
$E_B-\mu_B N=0$, i.e. the vacuum energy density $\rho_{\rm vac}$ in
the true vacuum B also becomes zero. 

We can readily apply this consideration to the quantum vacuum 
in our Universe. This condensed-matter example suggests that the
cosmological constant is zero before the cosmological phase transition.
During the non-equilibrium transient period of time, the microscopic
(Planckian) parameters of our vacuum are adjusted to a new equilibrium
state in a new vacuum, and after that the cosmological constant becomes
zero again.  Of course, we do not know what are these microscopic
parameters and how they relax in the new vacuum to establish the new
equilibrium. This already depends on the details of the system and cannot
be extracted from the analogy with quantum vacua in liquids. However,
using our experience with quantum liquids we can try to estimate the
range of change of the  microscopic parameters during the phase
transition.

Let us consider, for example, the electroweak phase transition, 
assuming that it is of the first order and thus can occur at low
temperature, so that we can discuss the transition in terms of the vacuum
energy. In this transition, the vacuum energy density changes from zero
in the initially equilibrium false vacuum to the negative value on the
order of 
\begin{equation}
\delta \rho_{\rm vac}^{\rm ew}\sim -\sqrt{-g}E_{\rm ew}^4 
\label{ElectroweakCorrection}
\end{equation}
in the true vacuum, where $E_{\rm ew}$ is the electroweak energy scale. 
 To restore the equilibrium, this negative energy must be compensated by
the adjustment of the microscopic (trans-Planckian) parameters. As such a
parameter we can use the value of Planck energy scale $E_{\rm Pl}$. It
determines the natural scale for the vacuum energy density $\sim
\sqrt{-g}E_{\rm Pl}^4$. This is the contribution to the vacuum energy
from the modes with the Planck energy scale. When the cosmological
constant is concerned, this contribution is effectively cancelled by the
microscopic (transplanckian) degrees of freedom in the equilibrium
vacuum, but otherwise it plays an important role in the energy balance and also in the quantum and thermodynamic fluctuations of the vacuum energy density about zero \cite{Fluctuations}.
Actually the same happens with the estimate in Eq. (\ref{3He}) of the
vacuum energy in quantum liquids: the  Eq. (\ref{3He}) highly
overestimates the magnitude of the cosmological constant, but it gives us
the correct estimate of the condensation energy, which is an important
part of the vacuum energy. 
  
Now, using the same argumentation as in quantum liquids, we can say that
the variation $\delta E_{\rm Pl}$ of this microscopic parameter $E_{\rm
Pl}$ leads to the following variation of the vacuum energy:
\begin{equation}
\delta \rho_{\rm vac}^{\rm Pl} 
\sim \sqrt{-g}E_{\rm Pl}^3 \delta E_{\rm Pl}~.
\label{Correction}
\end{equation}
In a new equilibrium vacuum, the density of the vacuum energy must be
zero  $\delta\rho_{\rm vac}^{\rm ew} +\delta\rho_{\rm vac}^{\rm Pl} =0$,
and thus the relative change of the microscopic parameter $E_{\rm Pl}$
which compensates the change of the electroweak energy after the
transition is

\begin{equation}
\frac{\delta E_{\rm Pl}}{E_{\rm Pl}} \sim 
\frac{E_{\rm ew}^4}{E_{\rm Pl}^4}~.
\label{Correction2}
\end{equation}
The response of the deep vacuum appears to be extremely small: 
the energy at the Planck scale is so high that a tiny variation of the
microscopic parameters is enough to restore the equilibrium violated by
the cosmological transition. The same actually occurs at the first-order
phase transition between $^3$He-A and $^3$He-B: the change in the energy 
of the superfluid  vacuum after transition is compensated by a tiny
change of the microscopic parameter -- the number density of $^3$He atoms
in the liquid: $\delta n/n \sim 10^{-6}$.

This remarkable fact may have some consequences for
 the dynamics of the cosmological constant after the phase transition.
Probably this implies that  $\lambda$ relaxes rapidly. But at the moment
we have no reliable theory describing the processes of relaxation of
$\lambda$ \cite{Alam,Padmanabhan2,Evolution}: the dynamics of $\lambda$
violates the Bianchi identity, and this requires the modification of the
Einstein equations. There are many ways of how to modify the Einstein
equations, and who knows, maybe the thermodynamic principles can show us
the correct one.

\section{Static Universes with and without gravity} 
\label{EinsteinUniverses}

As is well known there is a deep connection between Einstein's
 general relativity and the thermodynamic laws. It is especially spectacular
in application to the physics of the quantum vacuum in the presence of an
event horizon \cite{Bekenstein,HawkingNature} both in the
fundamental and induced gravity \cite{JacobsonInduced}. This connection
also allows us to obtain the equilibrium Einstein Universes from the
thermodynamic principles without solving the Einstein equations. Using
this derivation we can clarify how $\lambda$ responds to matter in
special and general relativity.

Let us start with the Universe without gravity, i.e. 
in the world  obeying the laws of special relativity. For the static
Universe, the relation between the matter and the vacuum energy is
obtained from a single condition:  the pressure in the equilibrium
Universe must be zero if there is no external environment, $p_{\rm
total}=p_{\rm matter}+ p_{\rm vac}=0$. This gives 
\begin{equation}
\rho_{\rm vac} =-p_{\rm vac} =p_{\rm matter}=
w_{\rm matter}\rho_{\rm matter}~,
\label{SpecialRelativity}
\end{equation}
where $p_{\rm matter}=w_{\rm matter}\rho_{\rm matter}$ 
is the equation of state for matter. In Sec.
\ref{non-gravitatingUniverse} this result was obtained for the
condensed-matter analogs of vacuum (superfluid condensate) and radiation
(gas of quasiparticles with
$w_{\rm matter}=1/3$).

 In the Universe obeying the laws of general relativity, 
the new player intervenes --  the gravitational field which contributes
to pressure and energy. But it also brings with it the additional
condition -- the gravineutrality, which states that the total energy
density in equilibrium Universe (including the energy of gravitational
field) must vanish, $\rho_{\rm total}=0$. This is the analog of the
electroneutrality condition, which states that both the spatially
homogeneous condensed matter and Universe must be electrically neutral,
otherwise due to the long-range forces the energy of the system is
diverging faster than the volume. In the same way the energy density,
which for the gravitational field plays the role of the density of the
electric charge, must be zero in equilibrium.  Actually the 
gravineutrality means the equation $\rho_{\rm total}+3p_{\rm total}=0$,
since
$\rho+3p$ serves as a source of the gravitational field in the
Newtonian limit, but we have already imposed the condition on pressure: $p_{\rm total}=0$. 
Thus we have two equilibrium conditions:
\begin{equation}
p_{\rm total}=p_{\rm matter}+ p_{\rm vac}+ p_{\rm gr}=0~,~\rho_{\rm total}=\rho_{\rm matter}+ \rho_{\rm vac}+ \rho_{\rm gr}=0 ~.
\label{CurvatureEnergy}
\end{equation}

As follows from the Einstein action for the gravitational field, 
the energy density of the gravitational field stored in the
spatial curvature is proportional  to
\begin{equation}
 \rho_{\rm gr}\propto  -{1\over GR^2}~.
\label{GrEnergy}
\end{equation}

Here $G$ is the Newton constant; $R$  the radius of the closed Universe; 
the exact factor is $-{3\over 8\pi}$, but this is not important for our
consideration.  The contribution of the gravitational field to pressure
is obtained from the conventional thermodynamic equation $p_{\rm
gr}=-d(\rho_{\rm gr}R^3)/d(R^3)$ . This gives the equation of state for
the energy and partial pressure induced by the gravitational field in the
Universe with a constant curvature:
\begin{equation}
 p_{\rm gr}=-{1\over 3}\rho_{\rm gr}~.
\label{CurvatureMass}
\end{equation}

This effect of the 3D curvature of the Universe can be compared to
 the effect of the 2D spatial curvature of the surface of a liquid drop.
Due to the surface tension the curved boundary of the liquid gives rise to
the Laplace pressure $p_\sigma=- 2\sigma/R$ , where $\sigma$ is the
surface tension. The corresponding energy density is the surface energy
divided by the volume of the droplet,
$\rho_\sigma=\sigma S/V= 3\sigma/R$ . This energy density and the Laplace
pressure obey the equation of state $p_\sigma=- (2/3)\rho_\sigma$ . If
there is no matter (quasiparticles), then the Laplace pressure must be
compensated by the positive vacuum pressure. As a result the negative
vacuum energy density arises in the quantum liquid when its vacuum is
disturbed by the curvature of the boundary:
$\rho_{\rm vac} =-p_{\rm vac} =p_\sigma=-2\sigma/R$ . This influence of
the boundaries on the vacuum energy  is the analog of
Casimir effect 
\cite{CasimirPaper} in quantum liquids. 

Returning to the Universe with matter and gravity, we must solve
the two equations (\ref{CurvatureEnergy}) by using the equations of state for
each of the three components $p_a=w_a\rho_a$, where  
$w_{\rm vac}=-1$ for the vacuum contribution;
$w_{\rm gr} =-1/3$ for the contribution of the gravitational field;
and $w_{\rm matter}$ for matter ($w_{\rm matter}=0$ for the cold matter 
and $w_{\rm matter}=1/3$ for the ultra-relativistic matter  and radiation
 field). 
The simplest solution of
these equations is, of course, the Universe 
without matter. This Universe is  flat, $1/R^2=0$, and the 
vacuum energy 
density in such a Universe is zero, $\lambda=0$. 
The vacuum is not perturbed, and thus its energy density is identically
zero.

The solution of the equations (\ref{CurvatureEnergy}) with matter gives
 the following value of the vacuum
energy  density in terms of matter density:
\begin{equation}
\rho_{\rm vac}= {1\over 2} \rho_{\rm matter} (1+3w_{\rm matter})~.
\label{EinsteinSolution}
\end{equation}
It does not depend on the Newton's constant $G$, and thus 
in principle it must be valid in the limit $G\rightarrow 0$. However, in
the world without gravity, i.e. in the world governed by Einstein's
special theory of relativity where $G=0$ exactly, the vacuum response to
matter in Eq. (\ref{SpecialRelativity}) is different. This demonstrates
that the special relativity is not the limiting case of general relativity.

In the considered simple case with three ingredients (vacuum, 
gravitational field, and matter of one kind) the two conditions
(\ref{CurvatureEnergy}) are enough to find the equilibrium configuration.
In a situation with more ingredients we can also use the thermodynamic
analysis, but now in terms of the free energy which must be minimized in order to
find the equilibrium Universe (see e.g. Ref. \cite{Barcelo}).
 
\section{Conclusion} 

The general thermodynamic analysis of the quantum vacuum,
 which is based on our knowledge of the vacua in condensed-matter
systems, is consistent with Einstein's earlier view on the cosmological
constant. In the equilibrium Universes the value of the cosmological
constant is regulated by matter. In the empty Universe, the vacuum energy
is exactly zero, $\lambda=0$. The huge contribution of the zero point
motion of the quantum fields to the vacuum energy is exactly cancelled by
the trans-Planckian degrees of freedom of the quantum vacuum without any
fine-tuning.  In the equilibrium Universes homogeneously filled with matter, the vacuum is disturbed, and the density of the vacuum energy
becomes proportional to the energy density of matter,  $\lambda=\rho_{\rm
vac}\sim \rho_{\rm matter}$. This takes place even within Einstein's
theory of special relativity, i.e. in a world without gravity, even though
the response of the vacuum to matter without gravity is different. 

So, instead of being "mein gr\"osster Fehler", $\lambda$ 
appeared to be one of the brilliant inventions of Einstein. It was
reinforced by  the quantum field theory and passed all the tests posed by
it. The thermodynamic laws hidden in Einstein's general theory of
relativity proved to be more general than relativistic quantum field
theory. Now we must move further -- out of the thermodynamic equilibrium.
Introducing $\lambda$, Einstein left us with the problem of how to relax
$\lambda$. This is a challenge for us  to find the principles which
govern the dynamics of $\lambda$. What can the quantum liquids, with
their quantum vacuum and effective QFT, say on that? Shall we listen to
them?

  I highly appreciate the hospitality extended to me during my visit to the
Scuola Normale Superiore in Pisa where this paper has been written.
  This work was also  supported  in part by the Russian Foundations for
Fundamental Research,  by the Russian Ministry of Education and Science
through the Leading Scientific School grant $\#$2338.2003.2 and through
the Research Programme "Cosmion", and by ESF COSLAB Programme.


\begin{thebibliography}{10}


\bibitem{einstein}
A.~Einstein, Kosmologische
Betrachtungen zur allgemeinen Relativit\"ats\-theorie,
Sitzungberichte der Preussischen Akademie der Wissenschaften,  {\bf 1},
142--152 (1917); also in a translated version in {\it The principle of
Relativity},  Dover (1952).

\bibitem{bib1} A. Einstein, Prinzipielles zur allgemeinen
Relativit\"atstheorie, Annalen der Physik
\textbf{55}, 241--244 (1918).


\bibitem{Weinberg2}  S. Weinberg,  The cosmological constant
problem,  Rev. Mod. Phys.  {\bf  61}, 1--23 (1989).

\bibitem{Padmanabhan}
 T. Padmanabhan, Cosmological constant - the weight of the vacuum,
  Phys. Rept. {\bf 380},  235--320 (2003)

\bibitem{Spergel} D.N. Spergel, L. Verde, H.V. Peiris,  {\it et al.}, 
First Year Wilkinson Microwave Anisotropy Probe (WMAP)
Observations: Determination of Cosmological Parameters,
Ap. J. Suppl. {\bf 148}, 175 (2003).

\bibitem{VolovikBook}
G.E. Volovik,
\emph{The Universe in a Helium Droplet}, Clarendon Press, Oxford (2003).

\bibitem{KlinkhamerVolovik} F.R. Klinkhamer and G.E. Volovik, 
 Emergent CPT violation from the
splitting of Fermi points, hep-th/0403037;
Quantum phase transition for
the BEC-BCS crossover in condensed matter physics and CPT violation in
elementary particle physics, JETP Lett.
{\bf 80}, (2004), cond-mat/0407597.

\bibitem{UnruhSonic} W.G. Unruh, Experimental black-hole
evaporation?, Phys. Rev. Lett. {\bf 46}, 1351--1354  (1981).    


\bibitem{AGDbook} A.A. Abrikosov, L.P. Gorkov   and I.E. Dzyaloshinskii,
\emph{Quantum Field Theoretical Methods in Statistical Physics},
Pergamon, Oxford (1965).

\bibitem{UniverseSpecialRelativity} 
G.E. Volovik, Vacuum energy and Universe in special relativity,
JETP Lett. {\bf 77}, 639--641 (2003), gr-qc/0304103.


\bibitem{Fluctuations} G.E. Volovik,  On thermodynamic and quantum fluctuations of
cosmological constant, Pisma ZhETF {\bf 80}, 531--534 (2004), gr-qc/0406005.

\bibitem{Alam} U. Alam, V. Sahni and A.A.
Starobinsky,  Is dark energy decaying?
JCAP 0304 002 (2003). 

\bibitem{Padmanabhan2} H.K. Jassal, J.S. Bagla and T. Padmanabhan, WMAP
constraints on low redshift evolution of dark energy,
astro-ph/0404378

\bibitem{Evolution}  G.E. Volovik,  Evolution of cosmological constant in
effective gravity, JETP Lett.  {\bf
77},  339-343 (2003), gr-qc/0302069; Phenomenology of effective gravity,
in: {\it Patterns of Symmetry Breaking}, H. Arodz et al. (eds.), Kluwer Academic
Publishers (2003), pp. 381--404, gr-qc/0304061.

\bibitem{Bekenstein} J.D. Bekenstein, Black holes and entropy, Phys. Rev. {\bf
D~7}, 2333 (1973).


\bibitem{HawkingNature} S.W. Hawking,  Black hole
explosions,  Nature {\bf 248},
30--31 (1974).


\bibitem{JacobsonInduced}  T. Jacobson, Black hole entropy and
induced gravity,  gr-qc/9404039.


\bibitem{CasimirPaper} H.G.B. Casimir, On the attraction between
two perfectly conducting plates \ldots ,  Kon. Ned.
Akad. Wetensch. Proc. {\bf 51}, 793 (1948).    



\bibitem{Barcelo} Carlos Barcel\'o and G.E. Volovik, 
A stable static Universe?  JETP Lett. {\bf 80}, 239--243 
(2004), gr-qc/0405105.



\end{thebibliography}
\end{document}